\documentclass[11pt,letterpaper]{article}
\usepackage[T1]{fontenc}
\usepackage{lmodern}
\usepackage[margin=1in]{geometry}
\usepackage{amsmath,amssymb,amsthm,mathtools}
\usepackage{microtype}
\usepackage{booktabs,array,enumitem}
\usepackage{tikz}
\usetikzlibrary{arrows.meta,decorations.pathreplacing,shapes.geometric,positioning}
\usepackage[numbers,sort&compress]{natbib}
\usepackage{thm-restate}

\makeatletter
\@ifundefined{newcounteralias}{}{%
  \renewcommand\thmt@autorefsetup{%
    \@xa\def\csname\thmt@envname autorefname\@xa\endcsname
      \@xa{\thmt@thmname}%
  }%
}
\makeatother

\usepackage[colorlinks=true,allcolors=blue!45!black]{hyperref}
\usepackage[nameinlink,noabbrev]{cleveref}
\setlist[itemize]{topsep=3pt,itemsep=1pt,parsep=0pt}
\setlist[enumerate]{topsep=3pt,itemsep=2pt,parsep=0pt}
\newtheorem{theorem}{Theorem}[section]
\newtheorem{lemma}[theorem]{Lemma}
\newcommand{\OPT}{\operatorname{OPT}}
\newcommand{\eps}{\varepsilon}
\title{A Constant-Factor Approximation for\\Capacitated Facility Location with Outliers}
\author{Rajni Dabas \\ 
\small rajni.dabas1@northwestern.edu \\
\small Northwestern University}
\date{}
\begin{document}
\maketitle
\begin{abstract}
We give a deterministic polynomial-time $(60/11+\eps)$-approximation for metric capacitated facility location with outliers, with arbitrary opening costs and nonuniform hard capacities. The algorithm respects both capacities and the outlier budget. Its guarantee, approximately $5.455$, is close to the best known factor $5+\eps$ for capacitated facility location without outliers. The main ingredient is a local-search comparison for a common rejection penalty that retains the signed difference between the algorithm's and the comparator's rejection counts. Ordinary approximation guarantees for the penalty objective do not provide this comparison. We extend the exchange-forest accounting of Bansal, Garg, and Gupta \cite{BGG18} so that a nonnegative assignment remainder pays for rejection and cancels in the final bound. We then merge two penalty assignments whose served counts bracket the outlier quota. Grouping alternating paths by their terminal facility preserves capacities and shares opening costs. These ingredients give a $(20/3+\eps)$-approximation; starting each penalty run from the preceding solution improves the guarantee to $60/11+\eps$.
\end{abstract}
\section{Introduction}\label{sec:intro}
In the classical facility location problem, we are given a set of clients
and a set of potential facilities in a metric space. Each facility has
an opening cost, and the cost of serving a client is its distance to the
facility serving it. The objective is to choose which facilities to open
and assign every client to an open facility so as to minimize the total
opening and service cost.

Several extensions of this problem have been studied. In
\emph{capacitated facility location} (CFL), each facility also has a
capacity that bounds the number of clients it can serve. We consider
hard capacities: a facility may be opened at most once, and its capacity
cannot be exceeded. These constraints make the problem substantially
more difficult. For example, the natural linear programming relaxation
has an unbounded integrality gap, even when capacities are
uniform~\cite{ASS14}. Local search has been particularly successful for
CFL. Starting with the work of P\'al, Tardos, and Wexler~\cite{PTW01},
successive improvements led to a $(5+\varepsilon)$-approximation by
Bansal, Garg, and Gupta~\cite{BGG12}.

Another extension allows a small number of clients to remain unserved.
Serving every client can be expensive when a few clients are far from
the facilities serving the rest. Allowing these clients to be omitted
can substantially reduce the cost. Charikar, Khuller, Mount, and
Narasimhan~\cite{CKMN01} introduced \emph{facility location with
outliers} (FLO), in which the input includes a budget $L$ on the number
of clients that may remain unserved. The objective is to minimize the
opening and service cost of serving the remaining clients. 

Combining these two extensions gives \emph{capacitated facility location
with outliers} (CFLO). A feasible solution must respect both the facility
capacities and the outlier budget. These requirements have been studied on its own. Their combination creates an additional difficulty: choosing the outliers changes where capacity is needed, while choosing facilities changes which clients are inexpensive to serve. An algorithm must therefore coordinate facility openings, assignments, and the choice of outliers.   

We study CFLO with arbitrary facility opening costs and nonuniform hard capacities. Our main result is a deterministic polynomial-time $(60/11+\eps)$-approximation that respects both capacities and the outlier budget. The factor is approximately $5.455$, close to the best known factor $5+\eps$ for capacitated facility location (with arbitrary opening costs) without outliers~\cite{BGG12,BGG18}. We first define the problem formally. %The main contribution is a penalty comparison that preserves the signed difference in rejection counts, together with a way to merge two penalty assignments while sharing facility opening costs.

\paragraph{Problem and result.}
Let $F$ be a set of $m$ facilities and $C$ a set of $n$ unit-demand clients in a finite metric space with distances $c$. Facility $i$ has opening cost $f_i\ge0$ and integer capacity $u_i\ge0$. An input quota $q\in\{0,\ldots,n\}$ specifies the minimum number of clients to serve; equivalently, at most $L=n-q$ clients may be outliers. A solution $S$ consists of open facilities $F_S\subseteq F$, served clients $C_S\subseteq C$, and an assignment $\sigma_S:C_S\to F_S$ with $|\sigma_S^{-1}(i)|\le u_i$ for every facility. Capacities are hard: each facility can be opened at most once. Let
\[
c_f(S)=\sum_{i\in F_S}f_i,\qquad
c_s(S)=\sum_{j\in C_S}c_{j,\sigma_S(j)},\qquad
c(S)=c_f(S)+c_s(S),
\]
and let $r(S)=n-|C_S|$. The objective is to minimize $c(S)$ subject to $|C_S|\ge q$. Denote the optimum by $\OPT$. Nonnegative costs allow an optimum serving exactly $q$ clients.

\begin{theorem}\label{thm:main}
For every $0<\eps\le1$, metric capacitated facility location with outliers admits a deterministic $(60/11+\eps)$-approximation. The running time is polynomial in the binary input length and $1/\eps$, for explicitly represented unit-demand clients.
\end{theorem}

The closest preceding exact-feasible result is the $(6.373+\eps)$-approximation of Gupta, Dabas, and Garg~\cite{GDG26}, which requires uniform facility opening costs and permits arbitrary hard capacities. More recently, Friggstad and Wu \cite{friggstad2026cflo-unif} improved the factor to $(3 + \epsilon)$ for the
same uniform-cost problem. Our result removes the uniform cost restriction. Arbitrary costs require controlling the cost of each newly opened facility: bounding only the number of openings cannot prevent repeated charges to an expensive facility.

\subsection{Outliers and capacity constraints in the literature}
The difficulty of adding outliers appears across clustering objectives. For capacitated $k$-center, Khuller and Sussmann~\cite{KS00} obtained factor $6$ for uniform hard capacities, and An et al.~\cite{ABCGMS15} obtained factor $9$ for arbitrary hard capacities. Cygan and Kociumaka~\cite{CK14} developed algorithms handling capacities and outliers simultaneously, with factors $23$ and $25$ for uniform and arbitrary hard capacities, respectively. Their results illustrate the substantial loss that can accompany enforcing both requirements in the same solution.

For $k$-median with outliers, Chen~\cite{Chen08} gave the first constant approximation respecting both the number of centers and the outlier budget. Krishnaswamy, Li, and Sandeep~\cite{KLS18} developed an iterative-rounding framework giving factor $7.081+\eps$, subsequently improved to $6.994+\eps$ by Gupta, Moseley, and Zhou~\cite{GMZ21}. Bicriteria algorithms provide another approach. Friggstad et al.~\cite{FKRS19} give a $(3+\epsilon)$ local-search guarantee that preserve the outlier budget while allowing $(1+\eps)k$ centers. They also show that bounded-size multiswap local search can have an unbounded locality gap when both bounds must be respected. This is an obstruction to that local-search neighborhood, rather than an inapproximability result for the problem. Dabas, Gupta, and Inamdar~\cite{DGI25} obtain a $(3+\eps)$-approximation for capacitated $k$-median with outliers in time fixed-parameter tractable in the number of centers, the number of outliers, and $1/\eps$.

There is also work on capacitated objectives with outliers for facility location. For uniform hard capacities and arbitrary opening costs, Dabas and Gupta~\cite{DG25} obtain a constant cost approximation allowing small violations of both capacities and the outlier budget; with uniform opening costs they eliminate the capacity violation. For uniform opening costs and hard capacities, Gupta, Dabas, and Garg~\cite{GDG26} obtained a $(6.373+\eps)$-approximation which was recently improved to $(3 + \epsilon)$ by Friggstad and Wu \cite{friggstad2026cflo-unif}. 

\subsection{Technical ideas and their relation to prior work}

We apply local search to a penalty variant of CFLO, in which instead of a hard bound on number of outliers, each unserved
client incurs a penalty and our goal now is to minimize the sum of facility costs, service costs and penalty costs. This problem is popularly known as {\em capacitated facility location with penalties} or the {\em capacitated prize-collecting facility location problem}. %We first define this variant formally and then explain how we use its solutions to satisfy the outlier budget.

%The input is as in CFLO, except that the requirement to serve at least $q$ clients is replaced by a nonnegative penalty $p_j$ for each unserved client $j$. We seek open facilities and a capacity-feasible assignment of any subset of clients minimizing$c(S)+\sum_{j\in C\setminus C_S}p_j$. There is no bound on the number of unserved clients. 

%We use the special case of a common penalty $p_j=\lambda\ge0$ for every client, whose objective is$\Phi_\lambda(S)=c(S)+\lambda\bigl(n-|C_S|\bigr)$. To recover a CFLO solution, we vary $\lambda$ and combine two penalty solutions whose numbers of served clients lie on opposite sides of $q$.The proof relies on the following three ingredients.

Our algorithm replaces the outlier quota by a common rejection penalty $\lambda\ge0$ and searches for improvements to
\[
\Phi_\lambda(S)=c(S)+\lambda r(S).
\]

It gradually increases $\lambda$, starting each run from the preceding solution, until the number of served clients reaches $q$. The proof has three ingredients.

\paragraph{A stronger guarantee than approximation of the penalty objective.}
Local search has already been studied for capacitated facility location with penalties~\cite{GG14}. Dabas and Gupta~\cite{DGCharging22} also give an approximation-preserving reduction from the capacitated prize-collecting problem to CFL, obtaining factor five by using CFL as a black box. Thus approximating the penalty objective is already possible at the CFL factor but this guarantee does not provide the comparison that we require. Against a quota optimum $O$ serving $q$ clients, it gives only
\[
c(S)+\lambda(q-|C_S|)
\le5\OPT+4\lambda(n-q).
\]
Averaging solutions to served count $q$ cancels the penalty term on the left, but leaves the uncontrolled term on the right. We instead prove
\begin{equation}\label{eq:intro-signed}
c(S)+3\lambda(q-|C_S|)\le5\OPT+\delta,
\end{equation}
where $\delta$ is a controlled local-search tolerance. The same coefficient multiplies the target count and the actual served count, so the penalty cancels at the required average. This signed comparison is the property that an ordinary penalty approximation cannot supply as a black box.

The exchange forest, its up/down organization, and its coefficient-three accounting come from Bansal, Garg, and Gupta~\cite{BGG18}. The obstacle introduced by rejection is that penalties need not be metric distances. Gupta and Gupta~\cite{GG14} already recognize this obstacle and use rejection as path endpoints. Our extension must additionally retain the signed count comparison through the forest analysis. We isolate a nonnegative remainder $D$ from paths ending at rejection. Metric transfers and rejections together cost at most $c_s(S)+c_s(O)+D$, while the assignment comparison contributes a negative multiple of $D$. The final bound leaves $-D$, so paying for rejection causes no loss in~\eqref{eq:intro-signed}. We also give polynomial procedures for searching the comparison assignments used by this analysis.

\paragraph{Merging assignments and sharing opening costs.}
Penalty search gives assignments $S^-$ and $S^+$ serving fewer than $q$ and at least $q$ clients, respectively. Lagrangian search and combinations of solutions have a long history in facility location~\cite{CKMN01,JV01}; our contribution is the particular integral merge needed for hard capacities and arbitrary opening costs. Representing facility capacities by labeled slots turns the two assignments into matchings. An augmenting path admits one additional client, preserves every previously served client, and can require opening only its terminal facility. Grouping paths by that facility lets many admissions share one opening charge. Selecting whole groups and at most one partial group produces a feasible solution within its fractional service budget, plus the cost of one additional facility. A maximum-opening-cost guess absorbs that charge. Combined with~\eqref{eq:intro-signed}, this already yields factor $20/3+\eps$.

\paragraph{Using descent across consecutive penalty runs.}
The two assignments contain more information than their separate approximation guarantees. Because the second run starts from the first, its potential at the new penalty cannot increase. Thus a large increase in ordinary cost must be supported by a correspondingly large increase in the number of clients served. The signed bound then forces the first solution to be inexpensive. Combining these two relations with the same merge improves the rounding loss from $4/3$ to $12/11$, giving $60/11+\eps$.

This separates the quota-recovery argument from the penalty analysis. Future improvements to CFL techniques could improve the CFLO guarantee if they can be extended to provide a suitable signed penalty comparison. An improved approximation for the ordinary CFL or penalty objective alone does not imply such a transfer; the signed inequality is the additional property that must be established.

\paragraph{Organization.}
%Local-search approximations for nonuniform hard-capacitated facility location originate with P\'al, Tardos, and Wexler~\cite{PTW01} and were improved by Zhang, Chen, and Ye~\cite{ZCY05} and Bansal, Garg, and Gupta~\cite{BGG12,BGG18}. An, Singh, and Svensson~\cite{ASS14} obtain a constant approximation from a stronger relaxation; the natural CFL relaxation has an unbounded integrality gap even with uniform capacities. 
Section~\ref{sec:framework} presents quota recovery and the warm-start refinement. Section~\ref{sec:penalty} describes the local-search algorithm, and Section~\ref{sec:signed} proves its signed guarantee. The appendices provide the complete merge proof, parameter choices, exchange construction, and search procedures.

\section{The approximation framework}
\label{sec:framework}
We first explain how the algorithm meets the service quota. The local-search algorithm used inside this framework is described in Section~\ref{sec:penalty}. Fix an instance with quota $q>0$, and let $O$ be an optimal solution
serving exactly $q$ clients, of cost $\OPT$. For a common penalty $\lambda\ge0$, the algorithm searches for improvements to $\Phi_\lambda(S)=c(S)+\lambda\bigl(n-|C_S|\bigr)$. Its guarantee is the following.

\begin{restatable}{theorem}{signedpenalty}\label{thm:signed}
For improvement threshold $\tau>0$, every terminal solution $S$ of the
penalty local search in satisfies
\begin{equation}\label{eq:signed-input}
 c(S)+3\lambda\bigl(q-|C_S|\bigr)\le5\OPT+6m\tau.
\end{equation}
\end{restatable}

The equal coefficients on $q$ and $|C_S|$ are essential. An ordinary
factor-five guarantee for $\Phi$ would give $c(S)+\lambda(n-|C_S|)\le5\OPT+5\lambda(n-q)$. Averaging solutions to served
count $q$ would leave the uncontrolled term $4\lambda(n-q)$. In contrast, the penalty term in~\eqref{eq:signed-input} cancels exactly at that average. The local search algorithm and the proof of Theorem \ref{thm:signed} appears in Sections \ref{sec:penalty} and \ref{sec:signed} respectively.

%The equal coefficients on $q$ and $|C_S|$ allow the penalty term tocancel when two solutions are averaged to served count $q$. An ordinary factor-five approximation for $\Phi_\lambda$ would instead leave an uncontrolled term $4\lambda(n-q)$.

\subsection{Combining two penalty solutions}
\label{sec:merge}
Increase the penalty on a grid of step $\eta>0$, starting each local-search run from the preceding solution. Stop at the first run whose solution serves at least $q$ clients. Let $S^-$ be the preceding solution and $S^+$ the current solution, obtained at penalties $\lambda^-$ and
$\lambda^+=\lambda^-+\eta$, respectively. Since we stop the first time a solution serves at least $q$ clients, the immediately preceding solution serves fewer than $q$ clients. Thus,
\[
 q^-=|C_{S^-}|<q\leq q^+=|C_{S^+}|.
\]

$S^-$ serves too few clients, while the cost of $S^+$ can be high, therefore, we try to combine the two solutions to obtain a solution of bounded cost serving at least $q$ clients and respecting the capacities. Let
\[
 a=c(S^-),\qquad b=c(S^+),\qquad
 \theta=\frac{q-q^-}{q^+-q^-},\qquad g=q^+-q^-.
\]
Thus $g>0$, $0<\theta\leq1$, and
$(1-\theta)q^-+\theta q^+=q$.
%Appendix~\ref{app:implementation} gives parameter choices that guarantee a crossing and polynomial running time.

%The first endpoint serves too few clients, whereas the second may beexpensive. We combine their assignments while sharing the opening costsof newly used facilities.

\begin{lemma}
\label{lem:merge}
Suppose $M\geq0$ bounds the opening cost of every facility open in $S^+$
and closed in $S^-$. In polynomial time, one can construct a
capacity-feasible solution serving at least $q$ clients and costing at most
\begin{equation}
  \min\{b,\ a+\theta b+M\}.
  \label{eq:merged-cost}
\end{equation}
The lemma requires nonnegative costs but does not require a metric.
\end{lemma}

\begin{proof}[Construction and proof outline]
Returning $S^+$ gives the first bound. For the second, represent every
facility by labeled capacity slots and view the assignments as matchings
between clients and slots. Their symmetric difference consists of
alternating paths and cycles. There are at least $g$ augmenting paths,
each of which admits one client to $S^-$ while preserving all clients
already served. Any subset can be switched simultaneously because the
paths are vertex-disjoint. Only a path's terminal facility may need to be
opened: every internal slot has an edge from the assignment of $S^-$.

Group these paths by terminal facility. Charge a group for its plus
edges, which bound the service-cost increase, and for its facility's
opening cost if necessary. The total charge of all groups is at most
$b$. Fractionally selecting $q-q^-$ paths therefore costs at most
$\theta b$. Ordering groups by charge per path gives a selection with
full groups and at most one partial group. In the partial group, take
the required number of paths with smallest service charge and pay its
entire opening cost. Service cost stays within its fractional budget,
and the extra opening charge is at most $M$. Keeping the facilities of
$S^-$ open gives cost at most $a+\theta b+M$. The full proof is in Appendix ~\ref{app:merge}.
\end{proof}

\subsection{A basic bound from penalty cancellation}
\label{sec:basic}
By the definition of $\theta$, we have,
$q-q_-=\theta g$ and $q-q_+=-(1-\theta)g$.
Applying~\eqref{eq:signed-input} to $S^-$ and $S^+$ gives
\[
  a+3\lambda^-\theta g\leq5\OPT+6m\tau,\qquad
  b-3\lambda^+(1-\theta)g\leq5\OPT+6m\tau.
\]
Multiply the first inequality by $1-\theta$ and the second by $\theta$.
Adding and using $\lambda^+-\lambda^-=\eta$, we obtain
\begin{equation}\label{eq:H}
 \begin{aligned}
 (1-\theta)a+\theta b
 &\le5\OPT+6m\tau+3\eta\theta(1-\theta)g\\
 &\le 5\OPT+6m\tau+3n\eta,
 \end{aligned}
\end{equation}
where the last inequality uses $g\le n$ and $\theta(1-\theta)\le1$.

Lemma~\ref{lem:merge} returns a feasible solution whose cost is at most
\begin{equation}\label{eq:coarse}
 \begin{aligned}
 \min\{b,a+\theta b+M\}
 & \le\frac{ (1-\theta)a+\theta b +(1-\theta)M}{1-\theta+\theta^2}\\
 &\le\frac43 (5\OPT+6m\tau+3n\eta)+M \qquad (\text{using }\ref{eq:H})\\
 &=\frac{20}{3}\OPT+8m\tau+4n\eta+M.
 \end{aligned}
\end{equation}

The second inequality uses
\[
 1-\theta+\theta^2=\frac34+\left(\theta-\frac12\right)^2\ge\frac34,
 \qquad
 \frac{1-\theta}{1-\theta+\theta^2}\le1,
\]

\paragraph{Absorbing the opening charge.}
Try every facility $i_0$ as a guess for a facility of maximum opening
cost in an optimal solution. In that branch, set $M=f_{i_0}$, discard
facilities costing more than $M$, and temporarily make $i_0$ free.
All remaining opening costs are at most $M$, so the merge lemma applies.
For a correct guess, an optimal original solution remains feasible
and its cost decreases by $M$. Hence the modified optimum satisfies
\[
 \OPT_0\le\OPT-M.
\]

Choose the improvement threshold and grid step so that
$8m\tau+4n\eta\le\varepsilon\OPT_0$. The merged solution then has modified cost at most
$(20/3+\varepsilon)\OPT_0+M$.
Restoring the guessed facility's opening cost adds at most another $M$,
so its original cost is at most
\[
 \begin{aligned}
 \left(\frac{20}{3}+\varepsilon\right)\OPT_0+2M
 &\le\left(\frac{20}{3}+\varepsilon\right)(\OPT-M)+2M\\
 &\le\left(\frac{20}{3}+\varepsilon\right)\OPT.
 \end{aligned}
\]
The last inequality uses $20/3+\varepsilon\ge2$ and $M\ge0$.
Trying all guesses and returning the cheapest feasible candidate,
measured under the original costs, therefore gives a
$(20/3+\varepsilon)$-approximation.

We next use the preceding penalty
run to obtain the advertised factor.

\subsection{The warm-start bound}
\label{sec:warm}

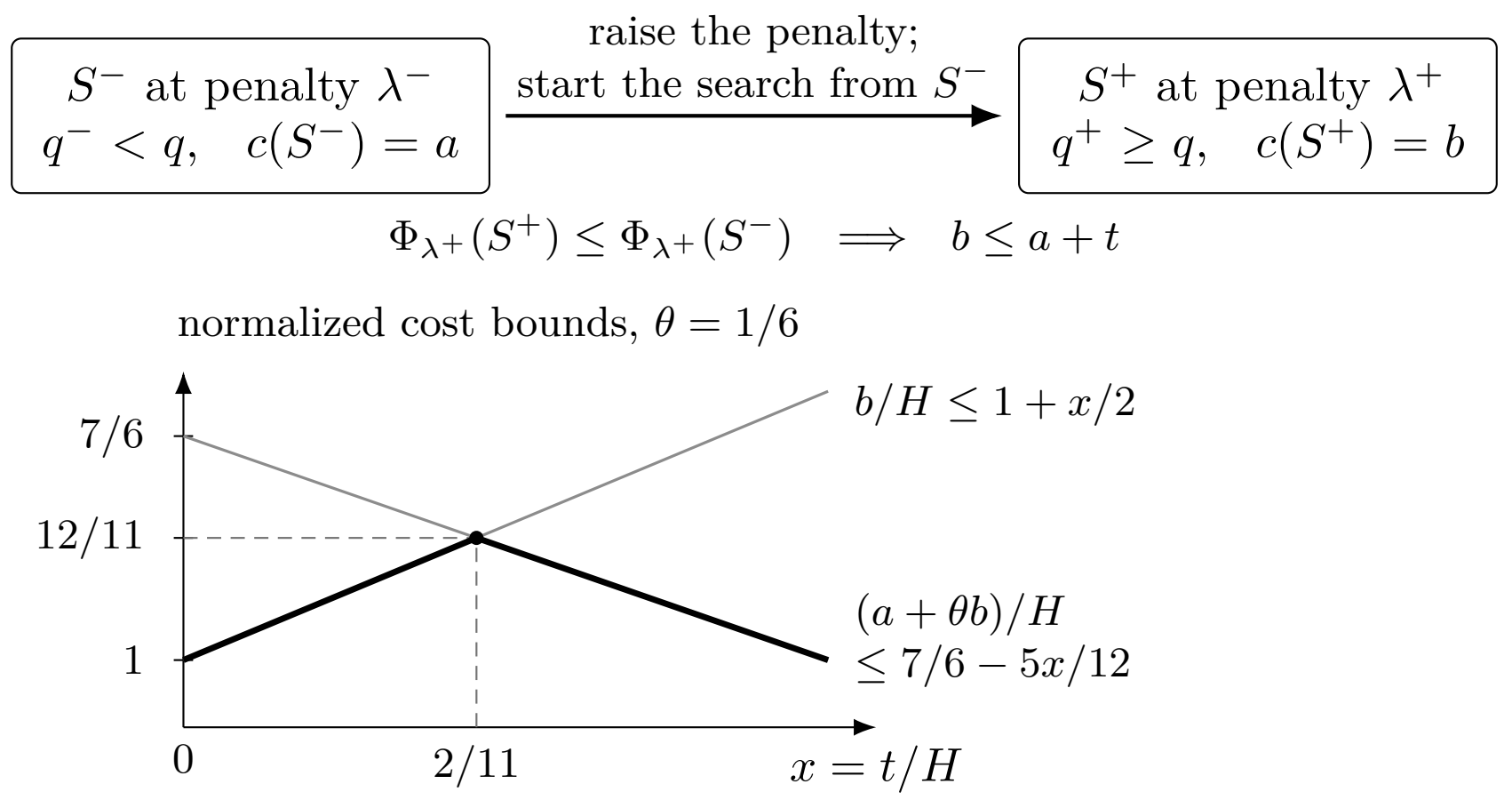
\begin{figure}[t]
\centering
\begin{tikzpicture}[font=\small,>=Latex]
  \node[draw,rounded corners=2pt,align=center,text width=3.3cm,minimum height=1.15cm]
    at (0,0) {$S^-$ at penalty $\lambda^-$\\$q^-<q$,\quad $c(S^-)=a$};
  \node[draw,rounded corners=2pt,align=center,text width=3.3cm,minimum height=1.15cm]
    at (7.5,0) {$S^+$ at penalty $\lambda^+$\\$q^+\ge q$,\quad $c(S^+)=b$};
  \draw[->,thick] (1.9,0)--(5.6,0)
    node[midway,above,align=center,font=\footnotesize]
      {raise the penalty;\\start the search from $S^-$};
  \node[align=center,font=\footnotesize] at (3.75,-.9)
    {$\Phi_{\lambda^+}(S^+)\le\Phi_{\lambda^+}(S^-)$\quad$\Longrightarrow$\quad$b\le a+t$};

  \begin{scope}[shift={(-.5,-14.05)},x=12cm,y=10cm]
    % The vertical axis is deliberately restricted to the relevant cost range.
    \draw[->] (0,.95)--(.43,.95) node[below] {$x=t/H$};
    \draw[->] (0,.95)--(0,1.215);
    \node[anchor=south west,font=\footnotesize] at (-.015,1.22)
      {normalized cost bounds, $\theta=1/6$};
    \foreach \yy/\lab in {1/1,1.090909/{12/11},1.166667/{7/6}} {
      \draw (-.006,\yy)--(.006,\yy);
      \node[anchor=east,font=\footnotesize] at (-.013,\yy) {$\lab$};
    }
    \node[below,font=\footnotesize] at (0,.95) {$0$};
    \draw[densely dashed,black!55] (0,1.090909)--(.181818,1.090909);
    \draw[densely dashed,black!55] (.181818,.95)--(.181818,1.090909);
    \node[below,font=\footnotesize] at (.181818,.95) {$2/11$};
    \draw[black!45,line width=.6pt] (0,1)--(.4,1.2);
    \draw[black!45,line width=.6pt] (0,1.166667)--(.4,1);
    \draw[black,line width=1.3pt] (0,1)--(.181818,1.090909)--(.4,1);
    \fill (.181818,1.090909) circle[radius=1.5pt];
    \node[anchor=west,font=\footnotesize] at (.405,1.19)
      {$b/H\le1+x/2$};
    \node[anchor=west,font=\footnotesize,align=left] at (.405,1.015)
      {$(a+\theta b)/H$\\$\displaystyle\le7/6-5x/12$};
  \end{scope}
\end{tikzpicture}
\caption{The additional information supplied by warm starting.
Here $t=\lambda^+(q^+-q^-)$, and $H$ includes the signed-comparison and
grid errors. Combining $b\le a+t$ with $a\le H-3\theta t$ gives an
increasing upper bound on $b$ and a decreasing upper bound on
$a+\theta b$. The thick curve is the smaller of the two bounds.
For the worst-case value $\theta=1/6$, its maximum is $12H/11$, attained
at $t/H=2/11$. In general the intersection value is
$H/(1-\theta+3\theta^2)\le12H/11$. The possible opening surcharge $M$
is added separately.}
\label{fig:warm}
\end{figure}

Define
\[
  H=5\OPT+6m\tau+3n\eta,\qquad t=\lambda^+g\geq0.
\]
The signed comparison for $S^-$, together with
$\lambda^-=\lambda^+-\eta$ and $\theta g\leq n$, gives
$a\leq H-3\theta t$.
The next search starts from $S^-$ and cannot increase its potential at
the new penalty.
Hence
\[
  b+\lambda^+(n-q^+)\leq a+\lambda^+(n-q^-),
\]
or $b\leq a+t$. The two endpoint bounds are therefore
\begin{equation}
  a\leq H-3\theta t,\qquad b\leq a+t.
  \label{eq:warm-endpoints}
\end{equation}
If $\theta\geq1/3$, they give
$b\leq H+(1-3\theta)t\leq H$, so $S^+$ suffices.
For $0<\theta<1/3$, the alternatives in the merge bound, before the
additional opening charge, satisfy
\[
  b\leq H+(1-3\theta)t,\qquad
  a+\theta b\leq(1+\theta)H-\theta(2+3\theta)t.
\]
Figure~\ref{fig:warm} illustrates these bounds for the worst-case value $\theta=1/6$. The first upper bound increases with $t$ and the second decreases.
They meet at $t=\theta H/(1-\theta+3\theta^2)$, where both equal
$H/(1-\theta+3\theta^2)$. On either side of the intersection, one bound
is no larger than this value. Since
\[
  1-\theta+3\theta^2
  =\frac{11}{12}+3\left(\theta-\frac16\right)^2\geq\frac{11}{12},
\]
Lemma~\ref{lem:merge} gives, in all cases, a feasible solution of cost
at most
\begin{equation}
  \min\{b,a+\theta b+M\}
  \leq\min\{b,a+\theta b\}+M
  \leq\frac{12}{11}H+M.
  \label{eq:warm-merge}
\end{equation}
The leading term is $(60/11)\OPT$. Guessing a maximum-opening-cost
facility of an optimum absorbs $M$; objective-scale guesses allow
$\tau$ and $\eta$ to be chosen without knowing $\OPT$.
Appendix~\ref{app:implementation} specifies the objective scales
and penalty parameters and completes the approximation and
running-time analysis.

% Requires tikz and \usetikzlibrary{arrows.meta}.

\section{The penalty local search}\label{sec:penalty}

Fix a common rejection penalty $\lambda\ge0$ and an improvement threshold
$\tau>0$.  For a current open set $F_S$, we always compute an assignment
minimizing
\[
  c_s(S)+\lambda r(S),\qquad r(S)=n-|C_S|,
\]
under the original capacities.  This is an integral minimum-cost flow
problem: a unit of flow for a client either uses an open facility or pays
$\lambda$ for rejection.  Consequently the open set determines the value
of the current solution, although the optimal assignment need not be
unique.  Write $w_i$ for its load at facility $i$, with $w_i=0$ if $i$ is
closed, and put $f(K)=\sum_{i\in K}f_i$.

\subsection{Operations and the values used to search them}

We use the standard \emph{add}, \emph{open}, and \emph{close} operations
of the local-search framework of Bansal, Garg, and Gupta~\cite{BGG18}.
Initially suppose that recipients are closed facilities.  An
$\operatorname{add}(t)$ operation opens a facility $t$.  An
$\operatorname{open}(t,K)$ operation opens $t$ and closes a set
$K\subseteq F_S$ whose total current load fits at $t$.
A $\operatorname{close}(s,T)$ operation closes one facility $s\in F_S$
and opens a set of recipients $T$; rejection remains allowed.
These descriptions specify changes to the open set.  After any change
we compute the \emph{best} penalty assignment under the original
capacities; the algorithm need not retain any particular routing used
to analyze the operation.

\paragraph{Already-open recipients.}
An open or close operation may also use an already-open recipient.
It has zero additional opening cost and only its currently unused
capacity is available for a transfer.  For uniform notation define
\[
 v_i=u_i-w_i,
 \qquad
 f_i'=\begin{cases}0,&i\in F_S,\\ f_i,&i\notin F_S.\end{cases}
\]
A recipient cannot also be a facility being closed.  Thus an open
operation has $K\subseteq F_S\setminus\{t\}$ and
$W_K:=\sum_{s\in K}w_s\le v_t$.  In a close operation the recipient
set is a subset of $F\setminus\{s\}$; only its closed members need to
be opened.  Closing $s$ saves its original opening cost $f_s$.

There are exponentially many possible sets $K$ and $T$.  We search
these operations through explicit, efficiently computable upper bounds
on their cost changes.  The following assignment-improvement values
make those bounds precise.

For a recipient $t$ and an integer $0\le b\le v_t$, let $A_t(b)$ be
the minimum change in $c_s+\lambda r$ when facilities in
$F_S\cup\{t\}$ are available, the load at $t$ is at most $w_t+b$,
and every other current facility $i$ has load at most $w_i$.
Opening costs are excluded.  Minimum-cost flow computes these values,
and optimality of the current assignment gives
\begin{equation}\label{eq:assignment-improvement}
 A_t(0)=0,\qquad A_t(b')\le A_t(b)\le0
 \quad(0\le b\le b'\le v_t).
\end{equation}
If $t$ is already open, all these values are zero: the restricted flow
region lies within the original feasible region and contains the
current assignment.

For an open operation we use the certificate
\begin{equation}\label{eq:open-certificate}
 \widehat\Delta_{\mathrm{open}}(t,K)
 = f_t'-f(K)+\sum_{s\in K}w_s c_{st}+A_t(v_t-W_K).
\end{equation}
For a close operation choose integers $0\le a_i\le v_i$ for its
recipients $i\in T$, with $\sum_{i\in T}a_i\le w_s$, and let
$h=w_s-\sum_{i\in T}a_i$.  Its certificate is
\begin{equation}\label{eq:close-certificate}
\begin{split}
 \widehat\Delta_{\mathrm{close}}(s,T,a)
 ={}&-f_s+\sum_{i\in T}f_i'+\sum_{i\in T}a_i c_{si}+\lambda h\\
 &+\min\bigl(\{0\}\cup\{A_t(v_t-a_t):t\in T\}\bigr).
\end{split}
\end{equation}
The minimum is zero when $T$ is empty.

Both certificates upper-bound the actual potential change.  To see
the role of $A_t$, first perform its assignment improvement while
reserving the places required by the transfer.  Move the clients
still at the closing facilities into those places, or reject the
allowed number.  Their remaining loads do not exceed their original
loads.  Moving a client from $s$ to $i$ increases its service cost by
at most $c_{si}$, by the triangle inequality, regardless of which
clients the assignment improvement leaves at $s$.  Rejecting a client
increases its penalized service cost by at most $\lambda$.
Appendix~\ref{app:search} justifies these certificates and gives
polynomial procedures for minimizing them. Only one $A_t$ term
is used in a close certificate; the analysis will not assume that
several independent assignment improvements can be realized together.

\subsection{The local-search procedure}

Given a starting solution $S_0$, the procedure
$\operatorname{Local}(\lambda,\tau;S_0)$ repeatedly does the following.
First it computes the optimal penalty assignment on the current open
set.  It then finds the best exact add change and the minimum open and
close certificates.  If all are at least $-\tau$, it returns the
current solution.  Otherwise it applies a change with value below
$-\tau$ and recomputes the optimal assignment under the original
capacities.

We call an output \emph{terminal}.  Every searched certificate is at
least $-\tau$ at a terminal solution, and every accepted change
decreases $\Phi_\lambda$ by more than $\tau$.  An exact add change is
computed by minimum-cost flow.  The certificate searches are polynomial
in $m,n$ and the binary encoding length, as proved in
Appendix~\ref{app:search}; the framework's implementation analysis bounds
the total number of accepted changes over all penalty runs.

\section{The signed penalty comparison}\label{sec:signed}

%\begin{theorem}\label{thm:signed}
%Let $O$ be an optimal quota solution serving exactly $q$ clients, and
%let $S$ be any terminal solution of the penalty local search at penalty
%$\lambda\ge0$ and threshold $\tau>0$.  Then
%\begin{equation}\label{eq:signed-comparison}
% c(S)+3\lambda\bigl(q-|C_S|\bigr)
% \le 5\OPT+6m\tau.
%\end{equation}
%\end{theorem}

\signedpenalty*

We prove the theorem by comparing assignments and then opening costs.
The exchange forest and its up/down organization come from
Bansal, Garg, and Gupta~\cite{BGG18}.  The additions needed here are to
keep rejection outside metric routes, retain a nonnegative remainder
that pays for rejection, and preserve the signs of the assignment
improvements when the exchange inequalities are summed.
Write $L=n-q$, so $r(O)=L$.

\subsection{Assignment paths and rejection}

\begin{figure}[t]
\centering
\begin{tikzpicture}[
  font=\small,>=Latex,
  real/.style={draw,circle,minimum size=6mm,inner sep=1pt},
  client/.style={draw,rectangle,minimum size=5mm,inner sep=1pt},
  receiving/.style={real,fill=black!15},
  edge/.style={->,line width=.65pt}
]
  \node[font=\footnotesize] at (3,1.02) {directed assignment path};
  \node[font=\footnotesize] at (8.1,1.02) {penalty change};
  \foreach \rr/\yy/\ss/\tt/\pen in
      {1/.4/s/t/0,2/-.4/\perp/t/{-\lambda},3/-1.2/s/\perp/{+\lambda},4/-2/\perp/\perp/0} {
    \ifnum\rr=2
      \node (s\rr) at (0,\yy) {$\ss$};
    \else\ifnum\rr=4
      \node (s\rr) at (0,\yy) {$\ss$};
    \else
      \node[real] (s\rr) at (0,\yy) {$\ss$};
    \fi\fi
    \node[client] (j\rr) at (1.15,\yy) {$j_0$};
    \node[real] (i\rr) at (2.3,\yy) {$i_1$};
    \node (d\rr) at (3.45,\yy) {$\cdots$};
    \node[client] (k\rr) at (4.6,\yy) {$j_\ell$};
    \ifnum\rr<3
      \node[receiving] (t\rr) at (5.85,\yy) {$\tt$};
    \else
      \node (t\rr) at (5.85,\yy) {$\tt$};
    \fi
    \draw[edge] (s\rr)--(j\rr);
    \draw[edge] (j\rr)--(i\rr);
    \draw[edge] (i\rr)--(d\rr);
    \draw[edge] (d\rr)--(k\rr);
    \draw[edge] (k\rr)--(t\rr);
    \node at (8.1,\yy) {$\pen$};
  }
\end{tikzpicture}
\caption{The four endpoint types in the assignment-path decomposition.
Arcs alternate between a client's assignment in $S$ and its assignment in
$O$; the omitted interior continues this alternation. Squares are clients,
circles are real facilities, and shaded circles are receiving endpoints.
Internal facilities are open in both solutions. Shifting a path changes
the rejection penalty only at its endpoints. The symbol $\perp$ denotes
rejection and has no metric distance.}
\label{fig:paths}
\end{figure}
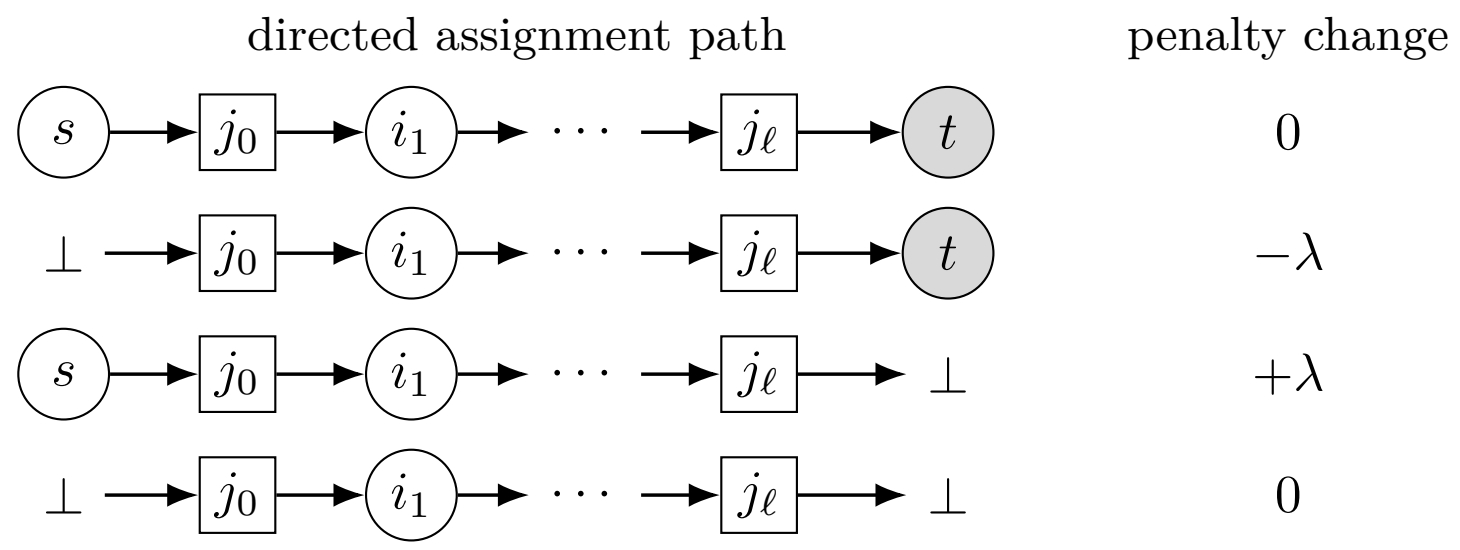

Give a rejected client the formal assignment $\perp$.  For each client
$j$, direct arcs from $\sigma_S(j)$ to $j$ and from $j$ to
$\sigma_O(j)$.  At each real facility pair incoming and outgoing arcs
as far as possible, and leave arcs at $\perp$ unpaired.  These pairs
decompose the arcs into paths and cycles, with every client in exactly
one component.  Internal real facilities belong to both open sets.
Maximal pairing leaves unmatched arcs in at most one direction at
each real facility. Figure~\ref{fig:paths} shows the four possible endpoint types and their penalty changes.

Let $S_j$ and $O_j$ be the service costs of $j$ in the two solutions,
using zero if it is rejected.  For a component $P$ set
\[
 \operatorname{length}(P)=\sum_{j\in P}(S_j+O_j),
 \qquad
 \operatorname{shift}(P)=\sum_{j\in P}(O_j-S_j).
\]
Shifting a path assigns each of its clients to its next facility.
Internal real loads do not change.  For a path from $s$ to $t$, the
change in service cost and rejection penalties is
\begin{equation}\label{eq:penalized-shift}
 \operatorname{shift}_\lambda(P)
 =\operatorname{shift}(P)
   +\lambda\bigl(\mathbf 1_{t=\perp}-\mathbf 1_{s=\perp}\bigr).
\end{equation}
Thus admission contributes $-\lambda$ and rejection contributes
$+\lambda$.  The symbol $\perp$ has no metric distance.

% Requires tikz and arrows.meta.

Let $\operatorname{in}(t)$ be the paths ending at a real facility $t$,
and let $\mathcal T$ contain those facilities with a nonempty such
set.  For $t\in\mathcal T$, define
\[
 \widehat u(t)=|\operatorname{in}(t)|
 =w_t(O)-w_t(S)\le v_t,
 \qquad
 G_t=\sum_{P\in\operatorname{in}(t)}
             \operatorname{shift}_\lambda(P).
\]
Every remaining path ends at $\perp$.  Its shift is feasible using
only current open facilities and increases no real load.  Optimality
of the current penalty assignment makes each such shift nonnegative.
Let $D\ge0$ be their total penalized shift.  Cycles have zero shift:
their shift is feasible for $S$, and the reverse shift is feasible
for $O$ and preserves its served count.  Optimality of the two
assignments gives the two opposite inequalities.  Summing over all
components therefore yields
\begin{equation}\label{eq:total-shift}
 \sum_{t\in\mathcal T}G_t+D
 =c_s(O)-c_s(S)-\lambda\bigl(r(S)-L\bigr).
\end{equation}
The groups $G_t$ include all incoming paths, including paths starting
at rejection or at a facility common to the two solutions.

We next extract transfer routes for the sources
$\mathcal S=F_S\setminus F_O$.  Every client at a source $s$ starts
a path.  Let $y_{st}$ count its paths to real target $t$, and let
$z_s$ count its paths to rejection.  Then
\[
 \sum_t y_{st}+z_s=w_s,
 \qquad Q_t:=\sum_{s\in\mathcal S}y_{st}\le\widehat u(t).
\]
For a real route, the triangle inequality gives
$c_{st}\le\operatorname{length}(P)$.  For a path from $s$ to
rejection,
\[
 \lambda\le
 \operatorname{shift}_\lambda(P)+\operatorname{length}(P).
\]
Its penalized shift is a nonnegative term of $D$.  The paths are
client-disjoint, so the total transfer and rejection charge satisfies
\begin{equation}\label{eq:transport-bound}
 \alpha:=\sum_{s,t}y_{st}c_{st}+\lambda\sum_s z_s
 \le c_s(S)+c_s(O)+D.
\end{equation}
This is the accounting that replaces a fictitious metric edge to
rejection.  Metric lengths pay for real routes; the same lengths
together with $D$ pay for rejected units.

\subsection{Savings in unused capacity}

Put $G_t^-:=\min\{G_t,0\}$.  If $b$ extra places are available at
$t$, shifting a fraction $x\le\min\{1,b/\widehat u(t)\}$ of every
incoming path is a feasible fractional assignment.  Every other real
facility has load at most its current load.  When $G_t<0$, the
penalized change is $xG_t$; when $G_t\ge0$, use the unchanged
assignment.  Minimum-cost-flow integrality turns the fractional
comparison into an integral assignment with no larger cost.
In particular, for an integer $0\le a\le\widehat u(t)$,
\begin{equation}\label{eq:certificate-bridge}
 A_t(v_t-a)
 \le\left(1-\frac{a}{\widehat u(t)}\right)G_t^-.
\end{equation}
Indeed, $v_t\ge\widehat u(t)$ leaves at least
$\widehat u(t)-a$ places for these shifts.  This inequality connects
the assignment-path analysis to the certificates actually searched
by the algorithm.

An exact add comparison, or assignment optimality when $t$ is already
open, gives
\begin{equation}\label{eq:add-comparison}
 f_t'+G_t^-\ge-\tau.
\end{equation}
For a closed $t$, the exact add change is at most
$f_t'+A_t(v_t)\le f_t'+G_t^-$, so terminality proves the claim.

Suppose an open comparison sends a group $K$ of total load
$W_K\le\widehat u(t)$ to $t$.  Apply
\eqref{eq:certificate-bridge} to
\eqref{eq:open-certificate}.  Since the searched certificate is at
least $-\tau$, we obtain
\begin{equation}\label{eq:open-comparison}
 f(K)\le f_t'+\sum_{s\in K}w_s c_{st}
       +\left(1-\frac{W_K}{\widehat u(t)}\right)G_t^-+\tau.
\end{equation}

The close comparison has an averaged form.  Give every formal
disposal recipient $d$ a positive integer analysis capacity
$\widehat u(d)$ and set $f_d'=G_d^-=0$.  Routing a unit to $d$ means
rejecting a client at charge $\lambda$; no metric distance is used.
If $R$ is a set of real or disposal recipients with positive analysis
capacities and integers $0\le x_{so}\le\widehat u(o)$ satisfy
$\sum_{o\in R}x_{so}=w_s$, then
\begin{equation}\label{eq:close-comparison}
\begin{split}
 f_s\le{}&\sum_{o\in R}f_o'
       +\operatorname{route}(s,x)
       +\sum_{o\in R}
           \left(1-\frac{x_{so}}{\widehat u(o)}\right)G_o^-\\
       &+\tau\bigl(1+|R\cap\mathcal T|\bigr).
\end{split}
\end{equation}
Here $\operatorname{route}(s,x)$ is the direct metric transfer charge
to real recipients plus the rejection charge.  To prove the
inequality, express the prescribed allocation as an average of
allocations with at most one partly filled recipient.  Each such
allocation uses at most one assignment improvement from
\eqref{eq:close-certificate}; add comparisons supply the terms for
unused real recipients.  Appendix~\ref{app:averaged-close} gives the
complete argument.  Thus \eqref{eq:close-comparison} does not require
independent improvements at several recipients in one operation.

\subsection{A bounded family of exchanges}

Cancel cycles in the real transport $y$ by alternating integral flow
adjustments of nonincreasing cost.  Its row sums, column sums $Q_t$,
and disposal quantities remain unchanged.  Attach a private disposal
leaf of quantity and analysis capacity $z_s$ to every source with
$z_s>0$.  The support is a forest, and
\eqref{eq:transport-bound} still holds.  Root each component containing
a real target at such a target.  A source with only disposal, or with
zero load, can be closed directly.

As in~\cite{BGG18}, call a nonroot node \emph{up} if its parent-edge
quantity is at least the sum of its child-edge quantities, and
\emph{down} otherwise; roots are down.  Write $K_u(i)$ and $K_d(i)$
for its up-children and down-children.  Up-source children fit into a
bounded number of open comparisons at their parent, with at most
one exceptional source closed separately.  A down-source first
routes to its children.  If its down-children have enough spare
allowance, they absorb its parent quantity.  Otherwise, ordering the
positive remainders increasingly allows each remainder to use the
next source's up-children; the final source uses its original row.
A disposal leaf is only an allowance for rejection.
Appendix~\ref{app:exchange} verifies the extension of the standard
forest construction to rejection and proves the following bounds.

\begin{lemma}\label{lem:exchange-counts}
There is a family of open and averaged close comparisons that closes
every source in $\mathcal S$ with total weight one.  Each real target
appears in at most five comparisons and receives total quantity at
most $2Q_t\le2\widehat u(t)$.  The total transfer and rejection
charge is at most $2\alpha$.  The total weight of the primitive open
and close comparisons is at most $m$.
\end{lemma}

For clarity, the recipient count has the following form; all
comparisons start from the same solution $S$.
\[
\begin{array}{c|c|c|c}
 \text{target}&\text{descendant comparisons}&
 \text{outside comparisons}&\text{total}\\ \hline
 \text{up}&\text{one open group; one final close}&
 \text{parent; predecessor}&\le4\\
 \text{down}&\text{two groups; exception; final close}&
 \text{parent only}&\le5
\end{array}
\]

\subsection{Summing the inequalities with their signs}

Let $a_t\le5$ count the uses of real target $t$, and let
$\beta_t\le2\widehat u(t)$ be the total quantity it receives.
In the sum of \eqref{eq:open-comparison} and
\eqref{eq:close-comparison}, the coefficients of $f_t'$ and $G_t^-$
are $a_t$ and $a_t-\beta_t/\widehat u(t)$, respectively.
Add $5-a_t$ copies of \eqref{eq:add-comparison}, including for
targets with $Q_t=0$.  The opening coefficient is now five and the
improvement coefficient is
\begin{equation}\label{eq:improvement-coefficient}
 \kappa_t=5-\frac{\beta_t}{\widehat u(t)}\ge3.
\end{equation}
Since $G_t^-\le0$, the direction of the resulting inequality is
\[
 \sum_t\kappa_tG_t^-
 \le3\sum_tG_t^-
 \le3\sum_tG_t.
\]
This is the step requiring sign control: a lower bound on a
coefficient would not yield this inequality for unrestricted signed
shifts.

The primitive comparisons contribute at most $m\tau$.  The extra add
tolerances in averaged closes and the final padding contribute at
most $5m\tau$: each real recipient occurrence in a close is already
included in its $a_t$ uses.  Lemma~\ref{lem:exchange-counts}, followed
by adding back facilities common to the two solutions, therefore gives
\begin{equation}\label{eq:opening-comparison}
 c_f(S)\le5c_f(O)+2\alpha+3\sum_{t\in\mathcal T}G_t+6m\tau.
\end{equation}
Substitute \eqref{eq:transport-bound} and \eqref{eq:total-shift}:
\begin{align*}
 c_f(S)
 &\le5c_f(O)+2\bigl(c_s(S)+c_s(O)+D\bigr)\\
 &\quad+3\bigl(c_s(O)-c_s(S)-\lambda(r(S)-L)-D\bigr)+6m\tau\\
 &=5c(O)-c_s(S)-3\lambda(r(S)-L)-D+6m\tau.
\end{align*}
Adding $c_s(S)$ and dropping the nonpositive remainder $-D$ proves
\eqref{eq:signed-input}, because
$r(S)-L=q-|C_S|$ and $c(O)=\OPT$.
The same remainder that paid for rejection in
\eqref{eq:transport-bound} thus appears with coefficient $2-3=-1$ in
the final comparison.\hfill$\square$

\appendix
\section{Proof of Lemma~\ref{lem:merge}} 
\label{app:merge}

Returning $S^+$ gives cost $b$. We construct the second solution
promised by the lemma, keeping the facilities of $S^-$ open and
admitting exactly $q-q^-$ additional clients.

Represent facility $i$ by $u_i$ labeled slots, each of capacity one.
Assign its clients to distinct slots in each solution. The two
assignments are matchings between clients and slots, with opening costs
still charged once per facility. Keep common matching edges fixed and
decompose the remaining edges into alternating paths and cycles. Call
an edge from $S^-$ a minus edge and one from $S^+$ a plus edge.

Let $\mathcal P$ be the paths with one more plus edge than minus edge,
and let $p=|\mathcal P|$. Each starts at a client unserved in $S^-$ and
ends at a slot unused in $S^-$. Switching a path deletes its minus
edges and inserts its plus edges. Every internal client remains served,
every internal slot remains occupied, and exactly one new client becomes
served. The paths are vertex-disjoint, so any subset can be switched
simultaneously without using a slot twice. Capacities remain feasible.
There are enough paths because
\[
  p\geq q^+-q^-=g:
\]
each path in $\mathcal P$ contributes $+1$ to the difference in matching
sizes, while every other component contributes $0$ or $-1$.

Every internal slot has a minus edge, so its facility is already open
in $S^-$. Only the terminal facility may need to be opened. Group the
paths by terminal facility. For a nonempty group $\mathcal P_i$, put
$p_i=|\mathcal P_i|$, and define
\[
  d(P)=\sum_{e\text{ a plus edge of }P}c_e,\qquad
  d_i=\sum_{P\in\mathcal P_i}d(P),\qquad
  f'_i=
  \begin{cases}
    0,&i\in F_{S^-},\\
    f_i,&i\notin F_{S^-}.
  \end{cases}
\]
Deleting minus edges only saves service cost, so $d(P)$ bounds the
service-cost increase from switching $P$. The paths have disjoint plus
edges, and every new terminal facility is open in $S^+$. Therefore
\[
  \sum_i d_i\leq c_s(S^+),\qquad
  \sum_i f'_i\leq c_f(S^+).
\]

Allow a fraction $x_i\in[0,1]$ of group $i$ to supply $x_ip_i$ paths
at charge $x_i(d_i+f'_i)$. Taking fraction $(q-q^-)/p$ of every group
supplies exactly $q-q^-$ paths at cost at most
\[
  \frac{q-q^-}{p}\sum_i(d_i+f'_i)
  \leq\frac{q-q^-}{g}\,b=\theta b.
\]
Order groups by $(d_i+f'_i)/p_i$ and take them in that order until the
required number of paths is reached, using only the needed fraction of
the last group. Exchanging fractional paths from a group with a higher
charge per path for paths from a cheaper group cannot increase cost.
Thus this choice minimizes fractional cost and has at most one partial
group. Its fractions satisfy
\[
  \sum_i x_ip_i=q-q^-,
  \qquad \sum_i x_i(d_i+f'_i)\leq\theta b.
\]

Switch every path in each full group. If there is a partial group
$\mathcal P_i$, the remaining number $h$ of paths is an integer because
$q-q^-$ and all full-group sizes are integers; also $x_i=h/p_i$.
Switch its $h$ paths with smallest $d(P)$. Their total service charge is
at most
\[
  \frac{h}{p_i}\sum_{P\in\mathcal P_i}d(P)=x_id_i.
\]
Thus the resulting service cost is at most
$c_s(S^-)+\sum_i x_id_i$.

Open each newly used terminal facility. Full groups pay exactly their
opening charge. In the sole partial group, paying the entire opening
cost instead of its fraction adds at most
$(1-x_i)f'_i\leq M$. Keeping all facilities of $S^-$ open, the total
cost is consequently at most
\[
  c(S^-)+\sum_i x_i(d_i+f'_i)+M
  \leq a+\theta b+M.
\]
Every client served in $S^-$ remains served, and exactly $q-q^-$ others
are admitted. Returning the cheaper of this solution and $S^+$ proves
Lemma~\ref{lem:merge}.

Capacities may be truncated to $n$, giving at most $mn$ slots. Finding
the matching decomposition, grouping its augmenting paths, sorting the
groups, and selecting paths from the partial group all take polynomial
time. This also proves the algorithmic assertion of the lemma.

%\input{fig_merge}

%\subsection{The elementary rounding inequality}

%Put $d=1-\theta+\theta^2$ and $W=(1-\theta)a+\theta b$.
%If $a\geq(1-\theta)b$, then $W\geq db$, so
%$\min\{b,a+\theta b\}\leq W/d$. If $a\leq(1-\theta)b$, then
%\[
 % W-d(a+\theta b)=\theta^2\bigl((1-\theta)b-a\bigr)\geq0,
%\]
%giving the same conclusion. Since
%$d=3/4+(\theta-1/2)^2\geq3/4$, the bound used in
%Section~\ref{sec:basic} follows:
%\[
 % \min\{b,a+\theta b+M\}
  %\leq\min\{b,a+\theta b\}+M
  %\leq\frac43 W+M.
%\]

\section{The complete quota algorithm and its running time}
\label{app:implementation}

We specify the parameters used in Section~\ref{sec:framework} and
prove polynomial running time. Assume $q>0$; otherwise return the empty
solution. Use the facility-cost branches of Section~\ref{sec:basic}:
for each facility $i_0$, set $M=f_{i_0}$, discard facilities costing more
than $M$, and make $i_0$ free. Skip infeasible branches. In each branch,
first test for a zero-cost solution by flow restricted to zero-cost
facilities and zero-cost connections. Record any such solution after
restoring the original opening costs, and skip that branch.

\paragraph{Scales and parameters.}
Consider a remaining branch with modified optimum $\OPT_0>0$.
Let $\ell$ be its smallest positive modified opening cost or retained
connection cost, and let $U$ be the cost of opening all retained facilities
and assigning $q$ clients by minimum-cost flow. Then
$\ell\leq\OPT_0\leq U$. Try $B=\ell2^j$ from $j=0$ through the first
scale at least $U$; one such scale satisfies
\begin{equation}
  \OPT_0\leq B\leq2\OPT_0.
  \label{eq:good-scale}
\end{equation}
For every scale, choose
\begin{equation}
  \tau=\frac{\eps B}{100m},\qquad
  \eta=\frac{\eps B}{100n},\qquad
  K=\left\lceil\frac{4B}{\eta}\right\rceil
   =\left\lceil\frac{400n}{\eps}\right\rceil,
  \label{eq:grid-parameters}
\end{equation}
where $m$ is the original number of facilities.
Starting with the empty solution $S_0$ at penalty zero, run
\[
  S_k=\operatorname{Local}(k\eta,\tau;S_{k-1}),
  \qquad k=1,\ldots,K,
\]
reoptimizing the starting assignment at each new penalty.
At the first $k$ with $|C_{S_k}|\geq q$, merge $S_{k-1}$ and $S_k$
using Lemma~\ref{lem:merge}, restore original opening costs, and record
the candidate. Discard scales with no crossing, and return the cheapest
recorded candidate under the original costs.

\paragraph{Approximation guarantee.}
Fix a branch and a scale satisfying~\eqref{eq:good-scale}.
If no crossing occurred, Theorem~\ref{thm:signed} and
$q-|C_{S_K}|\geq1$ would imply
\[
  3K\eta\leq5\OPT_0+6m\tau<12B\leq3K\eta,
\]
a contradiction. Thus consecutive solutions bracket the quota, and
their warm-start relation permits the bound
\eqref{eq:warm-merge}. Its error satisfies
\[
  \frac{12}{11}(6m\tau+3n\eta)
  \leq8m\tau+4n\eta
  =\frac{3\eps B}{25}
  \leq\eps\OPT_0.
\]
All retained opening costs are at most $M$, so the resulting modified
cost is at most $(60/11+\eps)\OPT_0+M$.
The opening-cost restoration argument in Section~\ref{sec:basic}
applies to any factor at least two, and therefore gives the claimed
$(60/11+\eps)$ bound for the correct facility guess. If that branch has
$\OPT_0=0$, its recorded zero-cost solution has restored cost at most
$M\leq\OPT$. The same error estimate also validates the basic bound
in Section~\ref{sec:basic}.

\paragraph{Running time.}
There are at most $m$ facility branches and
$O(1+\log(U/\ell))$ scales per branch. Each grid has
$K=O(n/\eps)$ points. Raising the penalty by $\eta$ increases the
preceding solution's potential by at most $n\eta$, whereas every
accepted operation decreases it by more than $\tau$. Since the initial
potential is zero and potentials remain nonnegative, telescoping over
the grid bounds the number $N$ of accepted operations by
\[
  N\tau\leq nK\eta,\qquad
  N\leq\frac{nK\eta}{\tau}=mK=O(mn/\eps).
\]
There is also at most one terminal search per grid point.
Assignment and certificate searches are polynomial by
Appendix~\ref{app:search}, as is the merge in Lemma~\ref{lem:merge}.
The number of scales and the encoding lengths of all generated rational
quantities are polynomial in the binary input length. Hence the entire
algorithm runs in time polynomial in the input length and $1/\eps$,
completing the proof of Theorem~\ref{thm:main}.
\section{The exchange forest in detail}\label{app:exchange}

We prove \eqref{eq:close-comparison} and
Lemma~\ref{lem:exchange-counts}, using the notation of
Section~\ref{sec:signed}. All comparisons start from the same terminal
solution $S$. We use the combinatorial forest construction of
Bansal, Garg, and Gupta~\cite[Sections~4--5]{BGG18}, and verify the
additional accounting needed for rejection.

\subsection{Averaging close comparisons}\label{app:averaged-close}

Fix a source $s$, a recipient set $R$ with positive integer analysis
capacities, and an integral allocation $x_{so}$ satisfying
$0\leq x_{so}\leq\widehat u(o)$ and $\sum_o x_{so}=w_s$.
For disposal recipients set $f_o'=G_o^-=0$ and charge $\lambda$ per unit.
Consider
\[
 \left\{x:0\leq x_o\leq\widehat u(o)\ (o\in R),\quad
                 \sum_{o\in R}x_o=w_s\right\}.
\]
Every vertex has at most one coordinate strictly between its bounds,
since two such coordinates could be perturbed in opposite directions.
The remaining coordinate is integral because the total and all bounds
are integral. Express the prescribed allocation as a convex combination
of these vertices.

At a vertex, close $s$, use its positive real recipients, and reject
the total allocated to disposal. If a real recipient $t$ is partial,
use the single assignment improvement supplied by
\eqref{eq:certificate-bridge}, of value at most
\[
 \left(1-\frac{x_t}{\widehat u(t)}\right)G_t^-.
\]
Otherwise use no improvement. This is a feasible comparison under
\eqref{eq:close-certificate}: the improvement reserves the required
places, and the remaining clients fit into these places and rejection
allowances.

For each unused real recipient add \eqref{eq:add-comparison}.
The coefficients of $f_o'$ and $G_o^-$ are then respectively one and
$1-x_o/\widehat u(o)$ for every real recipient, including full and unused
ones. Averaging preserves these affine coefficients and the routing
charge, proving \eqref{eq:close-comparison} with tolerance at most
$\tau(1+|R\cap\mathcal T|)$. The primitive close comparisons have total
weight one; each uses at most one assignment improvement.

\subsection{The forest with disposal leaves}

Use the cycle-free transport and rooting convention from
Section~\ref{sec:signed}. In the combinatorial construction
of~\cite{BGG18}, the source sizes are $w_s$, the real target sizes are
$Q_t=\sum_s y_{st}\leq\widehat u(t)$, and a private disposal leaf $d_s$
has size $z_s$. Thus incident quantities sum to the corresponding node
sizes. Roots are down and disposal leaves are up. Sources with zero
load or only disposal are closed directly.

The up-source grouping is exactly that of~\cite[Section~4.1]{BGG18}:
an up target uses one open comparison, while a down target uses at most
two, after removing the up-source child with largest parent-edge
quantity. Close this exceptional source separately using its original
forest row and \eqref{eq:close-comparison}.

For a down-source child $s$ of $t$, write
\[
 R(s)=y(s,t)-\sum_{o\in K_d(s)}y(s,o).
\]
If $R(s)\leq0$, send each child its original quantity and distribute the
parent quantity among down-children, adding at most $y(s,o)$ at each.
There is enough allowance by the definition of $R(s)$, and
$2y(s,o)\leq Q_o\leq\widehat u(o)$ for every down-child.
For the positive remainders, use the successor construction
of~\cite[Section~4.2]{BGG18}: order $0<R_1\leq\cdots\leq R_k$.
For $i<k$, send original quantities to up-children and doubled quantities
to down-children, and send the remaining $R_i$ units to up-children of
$s_{i+1}$. The final source uses its original row. Successor capacity
suffices because
\[
 R_i\leq R_{i+1}
 <\sum_{o\in K_u(s_{i+1})}y(s_{i+1},o).
\]
The forest ensures that a source's children and its successor's children
are disjoint. All quantities can be chosen integrally. A quantity sent
to a disposal leaf means rejection at cost $\lambda$ per unit, regardless
of its source. Apply \eqref{eq:close-comparison} to each close allocation.

% Requires tikz, shapes.geometric and decorations.pathreplacing.
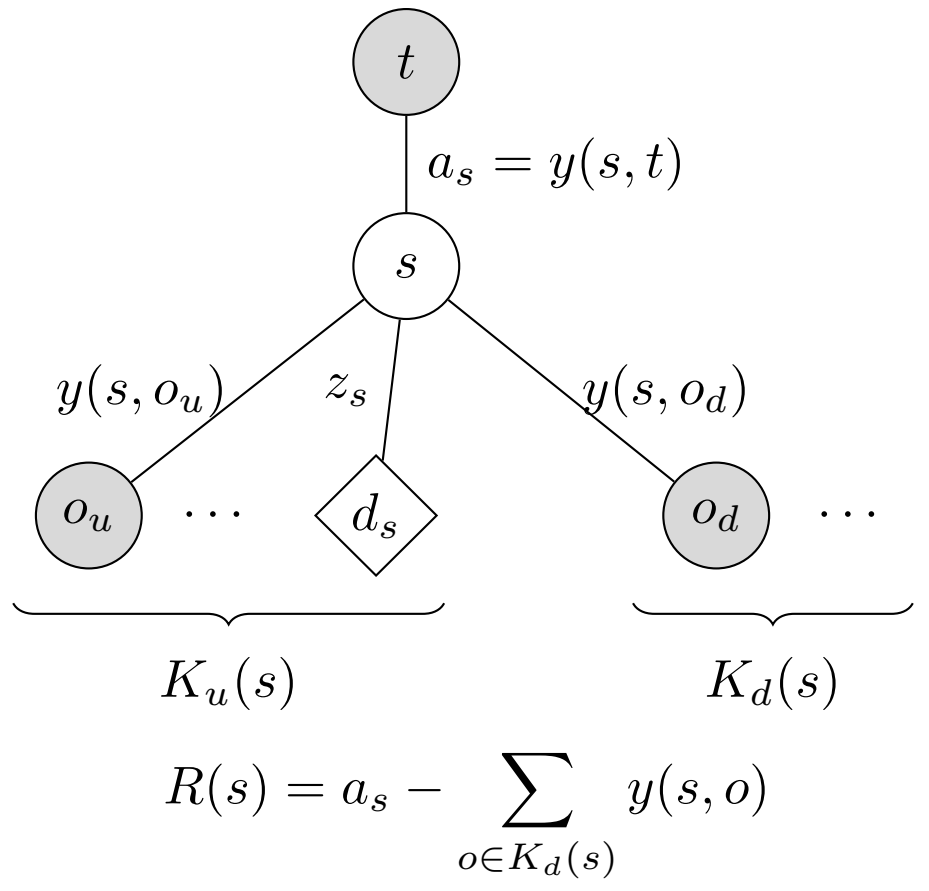
\begin{figure}[t]
\centering
\begin{tikzpicture}[
  font=\small,
  source/.style={draw,circle,minimum size=7mm,inner sep=1pt},
  target/.style={source,fill=black!15},
  disposal/.style={draw,diamond,aspect=1.1,minimum size=8mm,inner sep=1pt},
  brace/.style={decorate,decoration={brace,amplitude=4pt}}
]
  \node[target] (t) at (0,2.3) {$t$};
  \node[source] (s) at (0,.95) {$s$};
  \node[target] (u) at (-2.1,-.7) {$o_u$};
  \node (du) at (-1.25,-.7) {$\cdots$};
  \node[disposal] (d) at (-.2,-.7) {$d_s$};
  \node[target] (v) at (2.05,-.7) {$o_d$};
  \node (dv) at (2.95,-.7) {$\cdots$};
  \draw (t)--(s) node[midway,right] {$a_s=y(s,t)$};
  \draw (s)--(u) node[midway,left] {$y(s,o_u)$};
  \draw (s)--(d) node[midway,left] {$z_s$};
  \draw (s)--(v) node[midway,right] {$y(s,o_d)$};
  \draw[brace] (.25,-1.28)--(-2.6,-1.28);
  \node[anchor=north] at (-1.175,-1.5) {$K_u(s)$};
  \draw[brace] (3.35,-1.28)--(1.5,-1.28);
  \node[anchor=north] at (2.425,-1.5) {$K_d(s)$};
  \node[anchor=north] at (.4,-2.1)
    {$\displaystyle R(s)=a_s-\sum_{o\in K_d(s)}y(s,o)$};
\end{tikzpicture}
\caption{A down-source $s$ in the exchange forest. White circles are
sources and shaded circles are real targets. Its optional private disposal
leaf $d_s$ has quantity and analysis capacity $z_s$; it is always an
up-child and represents rejection at cost $\lambda$ per unit. The omitted
child targets may have further descendants. Down-child targets can absorb
up to one extra copy of their edge quantity. If $R(s)\le0$, they absorb
the parent quantity locally; if $R(s)>0$, the remaining quantity is sent
to a successor's up-children, except for the final source. Disposal is
charged directly and is never an intermediate point of a metric route.}
\label{fig:forest}
\end{figure}

\paragraph{Occurrences and quantities.}
The recipient counts are those of~\cite[Lemma~4]{BGG18}: an up target
has at most two descendant uses and two outside uses; a down target
has at most four descendant uses and one outside use. The
nonpositive-remainder case uses only children and creates no additional
parent or successor uses. The same edge-quantity count gives
\[
 a_t\leq5,\qquad \beta_t\leq2Q_t\leq2\widehat u(t).
\]
Each source is closed with total weight one. Each open group closes at
least one source, and each averaged close has primitive weight one, so
the total primitive-comparison weight is at most $m$.

\paragraph{Transfer and rejection charges.}
For real transfers, use the twice-per-edge charging
of~\cite[Sections~4.1--4.2, Lemma~3]{BGG18}. In particular, a successor
transfer to a real up-child $o$ is bounded by
\[
 c_{s_i o}\leq c_{s_i t}+c_{t s_{i+1}}+c_{s_{i+1}o}.
\]
The nonpositive-remainder case uses each child edge at most twice and
no parent edge, so it respects the same bound.
For disposal, charge $\lambda$ directly and use no metric route.
Each disposal allowance $z_s$ is used by its owner and at most one
predecessor, each to at most $z_s$, and hence contributes at most
$2\lambda z_s$. Disposal leaves are never intermediate metric vertices.
Consequently the total transfer and rejection charge is at most
\[
 2\left(\sum_{s,t}y_{st}c_{st}+\lambda\sum_s z_s\right)=2\alpha.
\]
Together with the occurrence and quantity bounds, this proves
Lemma~\ref{lem:exchange-counts}.
\section{Implementing the local-search certificates}\label{app:search}

We justify the certificates in Section~\ref{sec:penalty} and give
polynomial procedures for minimizing them. Truncate capacities to $n$,
since clients have explicitly represented unit demands.

\paragraph{Feasibility and assignment values.}
For either certificate, first perform its single auxiliary assignment
improvement, if any, while reserving the specified transfer capacity.
The loads remaining at facilities to be closed do not exceed their
original loads. Moving a client from $s$ to $i$ increases its service
cost by at most $c_{si}$, and rejection increases its penalized service
cost by at most $\lambda$. These bounds, together with the opening-cost
changes, prove \eqref{eq:open-certificate} and
\eqref{eq:close-certificate}. Unused transfer or rejection allowances
only reduce the charge. Reoptimizing the assignment after the open-set
change cannot increase its cost.

Compute $A_t(b)$ by minimum-cost flow on $F_S\cup\{t\}$ with
current-facility bounds $w_i$, recipient bound $b$ for closed $t$, and
rejection cost $\lambda$, subtracting the current penalized service cost.
Flow integrality gives integral assignments~\cite{AMO93}.
For already-open $t$, assignment optimality gives $A_t(b)=0$.
Assignment optimality and nested feasible regions give
\eqref{eq:assignment-improvement}.
All assignment tables require at most $m(n+1)$ flow computations;
exact add changes use further flows with the original capacities.

\paragraph{Open exchanges.}
Fix a recipient $t$. For each $s\in F_S\setminus\{t\}$, create a
knapsack item of weight $w_s$ and profit $f_s-w_sc_{st}$.
A standard $0$--$1$ knapsack dynamic program computes the maximum
profit $D(\ell)$ for each exact total load $0\leq\ell\leq v_t$.
The minimum open certificate is
\[
 f_t'+\min_{0\leq\ell\leq v_t}
          \{A_t(v_t-\ell)-D(\ell)\}.
\]
Backtracking recovers the closing set. This takes $O(mn)$ arithmetic
operations per recipient, or $O(m^2n)$ overall after computing the
assignment tables.

\paragraph{Close exchanges.}
Fix a source $s$ and an auxiliary recipient $t\ne s$.
Charge $f_t'$ separately, including when no client is transferred to
$t$, since it may be used solely for assignment improvement. Define,
for $i\ne s$ and $0\leq a\leq v_i$,
\[
 g_i(a)=
 \begin{cases}
  ac_{st}+A_t(v_t-a),&i=t,\\
  ac_{si}+\mathbf1_{a>0}f_i',&i\ne t.
 \end{cases}
\]
An allocation dynamic program computes
\[
 D(z)=\min\left\{\sum_{i\ne s}g_i(a_i):
          \sum_{i\ne s}a_i=z,\quad
          a_i\in\{0,\ldots,v_i\}\right\},
 \qquad 0\leq z\leq w_s.
\]
Process recipients one at a time, trying each allowable quantity at
each total-load state. The best value for this pair $(s,t)$ is
\[
 -f_s+f_t'+\min_{0\leq z\leq w_s}
                  \{D(z)+\lambda(w_s-z)\}.
\]
Also search the zero auxiliary option: use
$g_i(a)=ac_{si}+\mathbf1_{a>0}f_i'$ for every recipient and omit
the separate $f_t'$ charge. Backtracking recovers the allocation and
the recipients to open.

Every close certificate is represented by choosing a recipient
attaining its minimum auxiliary term, or the zero option; all other
unused recipients may be omitted without increasing its value.
Thus the search finds the minimum certificate without enumerating
recipient subsets. Each source--auxiliary-recipient pair takes
$O(mn^2)$ arithmetic operations, giving $O(m^3n^2)$ overall.
All load indices are at most $n$, so these procedures run in
polynomial time.
\section*{AI methodology}
The author had obtained the $(20/3+\eps)$-approximation before the warm-start refinement. GPT-6.1 Sol, with ultra reasoning, was used to obtain the warm-start improvement to $(60/11+\eps)$ and to improve the presentation of the paper.
\bibliographystyle{plainnat}
\bibliography{references}

@inproceedings{CK14,
  author = {Marek Cygan and Tomasz Kociumaka},
  title = {Constant Factor Approximation for Capacitated {$k$}-Center with Outliers},
  booktitle = {31st International Symposium on Theoretical Aspects of Computer Science},
  series = {LIPIcs}, volume = {25}, pages = {251--262}, year = {2014},
  doi = {10.4230/LIPIcs.STACS.2014.251}
}

@article{KS00,
  author = {Samir Khuller and Yoram J. Sussmann},
  title = {The Capacitated {$K$}-Center Problem},
  journal = {SIAM Journal on Discrete Mathematics},
  volume = {13}, number = {3}, pages = {403--418}, year = {2000},
  doi = {10.1137/S0895480197329776}
}

@article{ABCGMS15,
  author = {Hyung-Chan An and Aditya Bhaskara and Chandra Chekuri and Shalmoli Gupta and Vivek Madan and Ola Svensson},
  title = {Centrality of Trees for Capacitated {$k$}-Center},
  journal = {Mathematical Programming}, volume = {154}, number = {1--2},
  pages = {29--53}, year = {2015}, doi = {10.1007/s10107-014-0857-y}
}

@inproceedings{Chen08,
  author = {Ke Chen},
  title = {A Constant Factor Approximation Algorithm for {$k$}-Median Clustering with Outliers},
  booktitle = {Proceedings of the Nineteenth Annual ACM-SIAM Symposium on Discrete Algorithms},
  pages = {826--835}, year = {2008},
  url = {http://dl.acm.org/citation.cfm?id=1347082.1347173}
}

@inproceedings{KLS18,
  author = {Ravishankar Krishnaswamy and Shi Li and Sai Sandeep},
  title = {Constant Approximation for {$k$}-Median and {$k$}-Means with Outliers via Iterative Rounding},
  booktitle = {Proceedings of the 50th Annual ACM SIGACT Symposium on Theory of Computing},
  pages = {646--659}, year = {2018}, doi = {10.1145/3188745.3188882}
}

@article{FKRS19,
  author = {Zachary Friggstad and Kamyar Khodamoradi and Mohsen Rezapour and Mohammad R. Salavatipour},
  title = {Approximation Schemes for Clustering with Outliers},
  journal = {ACM Transactions on Algorithms}, volume = {15}, number = {2},
  articleno = {26}, numpages = {26}, year = {2019}, doi = {10.1145/3301446}
}

@inproceedings{GMZ21,
  author = {Anupam Gupta and Benjamin Moseley and Rudy Zhou},
  title = {Structural Iterative Rounding for Generalized {$k$}-Median Problems},
  booktitle = {48th International Colloquium on Automata, Languages, and Programming},
  series = {LIPIcs}, volume = {198}, pages = {77:1--77:18}, year = {2021},
  doi = {10.4230/LIPIcs.ICALP.2021.77}
}

@article{DG25,
  author = {Rajni Dabas and Neelima Gupta},
  title = {Uniform Capacitated Facility Location with Outliers/Penalties},
  journal = {Discrete Optimization}, volume = {55}, pages = {100878},
  year = {2025}, doi = {10.1016/j.disopt.2025.100878}
}

@article{DGI25,
  author = {Rajni Dabas and Neelima Gupta and Tanmay Inamdar},
  title = {{FPT} Approximation for Capacitated Clustering with Outliers},
  journal = {Theoretical Computer Science}, volume = {1027}, pages = {115026},
  year = {2025}, doi = {10.1016/j.tcs.2024.115026}
}

@inproceedings{GDG26,
  author       = {Rajni Dabas and
                  Naveen Garg and
                  Neelima Gupta},
  editor       = {Jens Vygen and
                  Jaroslaw Byrka},
  title        = {Capacitated Facility Location with Outliers and Uniform Facility Costs},
  booktitle    = {Integer Programming and Combinatorial Optimization - 25th International
                  Conference, {IPCO} 2024, Wroc{\l}aw, Poland, July 3-5, 2024, Proceedings},
  series       = {Lecture Notes in Computer Science},
  volume       = {14679},
  pages        = {85--98},
  publisher    = {Springer},
  year         = {2024},
  url          = {https://doi.org/10.1007/978-3-031-59835-7\_7},
  doi          = {10.1007/978-3-031-59835-7\_7},
  bibsource    = {dblp computer science bibliography, https://dblp.org}
}

@inproceedings{PTW01,
  author = {Martin P{\'a}l and {\'E}va Tardos and Tom Wexler},
  title = {Facility Location with Nonuniform Hard Capacities},
  booktitle = {42nd IEEE Symposium on Foundations of Computer Science},
  pages = {329--338}, year = {2001}, doi = {10.1109/SFCS.2001.959907}
}

@inproceedings{BGG12,
  author = {Manisha Bansal and Naveen Garg and Neelima Gupta},
  title = {A 5-Approximation for Capacitated Facility Location},
  booktitle = {Algorithms -- ESA 2012},
  series = {Lecture Notes in Computer Science}, volume = {7501},
  pages = {133--144}, publisher = {Springer}, year = {2012},
  doi = {10.1007/978-3-642-33090-2_13}
}

@inproceedings{BGG18,
  author = {Manisha Bansal and Naveen Garg and Neelima Gupta},
  title = {A 5-Approximation for Universal Facility Location},
  booktitle = {38th IARCS Annual Conference on Foundations of Software Technology and Theoretical Computer Science},
  series = {LIPIcs}, volume = {122}, pages = {24:1--24:12},
  publisher = {Schloss Dagstuhl -- Leibniz-Zentrum f{\"u}r Informatik},
  year = {2018}, doi = {10.4230/LIPIcs.FSTTCS.2018.24}
}

@inproceedings{ASS14,
  author = {Hyung-Chan An and Mohit Singh and Ola Svensson},
  title = {{LP}-Based Algorithms for Capacitated Facility Location},
  booktitle = {55th IEEE Annual Symposium on Foundations of Computer Science},
  pages = {256--265}, publisher = {IEEE Computer Society},
  year = {2014}, doi = {10.1109/FOCS.2014.35}
}

@book{AMO93,
  author = {Ravindra K. Ahuja and Thomas L. Magnanti and James B. Orlin},
  title = {Network Flows: Theory, Algorithms, and Applications},
  publisher = {Prentice Hall}, year = {1993}
}

@inproceedings{CKMN01,
  author = {Moses Charikar and Samir Khuller and David M. Mount and Giri Narasimhan},
  title = {Algorithms for Facility Location Problems with Outliers},
  booktitle = {Proceedings of the Twelfth Annual ACM-SIAM Symposium on Discrete Algorithms},
  pages = {642--651}, year = {2001},
  url = {https://www.cs.umd.edu/users/samir/grant/outlier.pdf}
}

@misc{GG14,
  author = {Neelima Gupta and Shubham Gupta},
  title = {Approximation Algorithms for Capacitated Facility Location Problem with Penalties},
  year = {2014}, eprint = {1408.4944}, archivePrefix = {arXiv},
  primaryClass = {cs.DS}, url = {https://arxiv.org/abs/1408.4944}
}

@article{JV01,
  author = {Kamal Jain and Vijay V. Vazirani},
  title = {Approximation Algorithms for Metric Facility Location and {$k$}-Median Problems Using the Primal-Dual Schema and Lagrangian Relaxation},
  journal = {Journal of the ACM}, volume = {48}, number = {2},
  pages = {274--296}, year = {2001}, doi = {10.1145/375827.375845}
}

@misc{DGCharging22,
  author = {Rajni Dabas and Neelima Gupta},
  title = {Locating Charging Stations: Connected, Capacitated and Prize-Collecting},
  year = {2022}, eprint = {2202.06262}, archivePrefix = {arXiv},
  primaryClass = {cs.DS}, url = {https://arxiv.org/abs/2202.06262}
}

@misc{friggstad2026cflo-unif,
      title={Multiple Swap Local Search Approximations for Uniform-Cost Capacitated Facility Location with Extensions to Outliers}, 
      author={Zachary Friggstad and Pengfei Wu},
      year={2026},
      eprint={2610.41187},
      archivePrefix={arXiv},
      primaryClass={cs.DS},
      url={https://arxiv.org/abs/2610.41187}, 
}
\end{document}